\documentstyle[preprint,aps,pre]{revtex}
\tighten

\begin{document}

\title{Effect of Non Gaussian Noises on the Stochastic
Resonance-Like Phenomenon in Gated Traps}

\author{Jorge A. Revelli $^1$\thanks{Fellow CONICET, Argentina;
e-mail: revelli@cab.cnea.gov.ar} Alejandro D. S\'anchez $^2$
\thanks{E-mail: asanchez@ifca.unican.es}, and Horacio S. Wio
$^1$ \thanks{Member of CONICET, Argentina; e-mail:
wio@cab.cnea.gov.ar}}

\address{1) Grupo de F\'{\i}sica Estad\'{\i}stica
\thanks{http://www.cab.cnea.gov.ar/CAB/invbasica/FisEstad/estadis.htm}\\
Centro At\'omico Bariloche (CNEA) and Instituto Balseiro (CNEA and
UNC) \\ 8400 San Carlos de Bariloche, Argentina\\ 2) Instituto de
F\'{\i}sica de Cantabria, Facultad de Ciencias,\\ Universidad de
Cantabria, Santander, Spain}

\maketitle

\begin{abstract}
We exploit a simple one-dimensional trapping model introduced
before, prompted by the problem of ion current across a biological
membrane. The voltage-sensitive channels are open or closed
depending on the value taken by an external potential that has
two contributions: a deterministic periodic and a stochastic one.
Here we assume that the noise source is colored and non Gaussian,
with a $q$-dependent probability distribution (where $q$ is a
parameter indicating the departure from Gaussianity). We analyze
the behavior of the oscillation amplitude as a function of both
$q$ and the noise correlation time. The main result is that in
addition to the resonant-like maximum as a function of the noise
intensity, there is a new resonant maximum as a function of the
parameter $q$.

\end{abstract}

\vskip 1.truecm

\normalsize

The growing interest in stochastic resonance (SR) has motivated a
wealth of studies in physical, chemical and biological systems
\cite{SR3}. In particular, it has been found to play a relevant
role in several problems in biology: mammalian sensory systems,
increment of tactile capacity, visual perception, low frequency
effects and low amplitude electromagnetic fields, etc
\cite{biology}.

In particular, there is an experiment on  SR related to the
measurement of the current through voltage-sensitive ion channels
in a cell membrane \cite{nature}. These channels switch (randomly)
between open and closed states, thus controlling the ion current.
This and other related phenomena have stimulated several
theoretical studies of the problem of ionic transport through
biomembranes, using different approaches, as well as different
ways of characterizing stochastic resonance in such systems
\cite{membrana}. It is worth remarking here that recent detailed
studies on the source of fluctuations in some biological systems
\cite{caltech} clearly indicate that noise sources in general
could be non Gaussian and that their distribution bounded.

In Ref. \cite{we1} we have studied a toy model, prompted by the
work in \cite{nature}, sketching the behaviour of an ion channel.
This included the simultaneous action of a deterministic and a
stochastic external field on the trapping rate of a gated
imperfect trap. Rather than attempting a precise modeling of the
behaviour of an ionic channel, we proposed there a simple model of
dynamical trap behaviour. Our main result was that even such a
simple model of a gated trapping process shows a SR-like
behaviour. In that initial study we assumed that the stochastic
external field was a Gaussian white noise.

In this work, and due to recent experimental evidences of the
boundness and non Gaussian character of noise sources in
biological systems \cite{caltech}, we analyze the same model, but
using a correlated non Gaussian noise source. Here we have
Gaussian or non Gaussian behaviour of the noise probability
distribution (pdf) depending on whether a parameter $q$ is $=1$
or $\neq 1$, respectively. If $q > 1$, the pdf results more
extended than a Gaussian one, while in the other case, that is $q
< 1$, the pdf is bounded. This latter aspect is of overwhelming
relevance for the problem of ion currents through membranes.

The study of gated trapping processes, i.e. a trapping process
where the traps have some kind of internal dynamic has attracted
considerable interest \cite{varios}. Many authors discussed the
way to link the gated trapping processes with the measured
behavior of the so called ionic pumps \cite{membrana}. For
example, among other factors, the ion transport depends on the
membrane electric potential (which plays the role of the barrier
height) and can be stimulated by both {\it dc} and {\it ac}
external fields.

In Ref.\cite{we1}, the study was based on the so called {\it
stochastic model} for reactions \cite{stoch,stoch1,stoch2},
generalized in order to include the internal dynamics of traps.
The dynamical process consists of the opening or closing of the
traps according to an external field. Such a field has two
contributions, one periodic with a small amplitude, and the other
stochastic whose intensity will be (as usual) the tuning
parameter. The starting model equation was
\begin{equation}
\label{model}
\partial_t \rho(x,t)=D \partial^2_x \rho(x,t)- \gamma(t) \delta(x)
\rho(x,t)+n_u,
\end{equation}
where $\gamma$ is a stochastic process that represents the
absorption probability of the trap, $\rho$ is the particle density
(particles that have not been yet trapped); for a given
realization of $\gamma$, $x$ is the coordinate over the
one-dimensional system and $n_u$ is a source term that represents
a constant flux of ions. The injection of ions can be at a trap
position or at any other position. In this last case the ion can
diffuse to the trap position. This diffusion coefficient would
represent an effective diffusion through the volume rather a
diffusion over the membrane surface.

The absorption is modelled as
\begin{equation}
\label{gamma} \gamma(t)=\gamma^* \theta[B \sin (\omega t) + \xi -
\xi_c],
\end{equation}
where $\theta (x)$, the Heaviside function, determines when the
trap is open or closed. The trap works as follows: if the signal,
composed of the harmonic part plus $\xi$ (the noise contribution),
reaches a threshold $\xi_c$ the trap opens, otherwise it is
closed.  We are interested in the case where $\xi_c>B$, that is,
without noise the trap is always closed. When the trap is open
the particles are trapped  with a given frequency (probability
per unit time) $\gamma^*$. In other words the open trap is
represented by  an ``imperfect trap" .  Finally, in order to
complete the model, we must give the statistical properties of
the noise $\xi$. In \cite{we1} we assumed that $\xi$ is an
uncorrelated Gaussian noise of intensity $\xi_0$. Here we use a
``colored" non Gaussian noise given by
\begin{equation}
\dot{\xi} = -\frac{1}{\tau }\frac{d}{d{\xi }}V_{q}(\xi )
+\frac{1}{\tau }\eta (t)  \label{nu}
\end{equation}
where $\eta (t)$ is a Gaussian white noise of zero mean and
correlation $<\eta (t)\eta (t^{\prime })>= 2 D \delta (t-t^{\prime
})$, and $V_{q}(\eta )$ is given by \cite{borland1}
\begin{equation}
V_{q}(\xi )=\frac{1}{\beta (q-1)}\ln [1+\beta (q-1)\frac{\xi^{2}
}{2}],
\end{equation}
where $\beta =\tau /D$. When $q \to 1$ we recover the
limit of $\xi $ being an Ornstein-Uhlenbeck process.

We define the current through the trap as ${\it J}(t)=\langle
\gamma_j(t) \rho(j l,t) \rangle$. The brackets mean averages over
all realizations of the noise. In \cite{we1}, that is in the case
of $\xi (t)$ being a Gaussian white noise, we have obtained some
analytical results and solved the equation numerically. However,
here we should resort only to Monte Carlo simulations.

As in \cite{we1}, we choose to quantify the SR-like phenomenon by
computing the amplitude of the oscillating part of the absorption
current given by ${\Delta \it J}={\it J}|_{\sin(\omega t)=1} -{\it
J}|_{\sin(\omega t)=-1}$. The qualitative behaviour of the system
can be explained as follows. For small noise intensities the
current is low (remember that $\xi_c>B$), hence ${\Delta \it J}$
is small too. For a large noise intensity the deterministic
(harmonic) part of the signal becomes irrelevant and the ${\Delta
\it J}$ is also small. Therefore, there must be a maximum at some
intermediate value of the noise.

The simulations were performed on a one dimensional lattice of $L$
sites with periodic boundary conditions. Initially there are no
particles on the lattice. The particles are injected randomly
every $1/(n_u L)$ units of time, with uniform distribution over
the lattice, and are allowed to perform a continuous time random
walk (characterized by the jump frequency at each neighbor site).
There is no restriction on the number of particles at each site.
A particle can be removed from the system with a given
probability distribution characterized by $\gamma$ when it
reaches the trap site. A detailed description of the algorithm
used can be found in \cite{stoch2}. The reaction times were
generated according to the following probability density function
\begin{equation}
p(t)=\exp  \left( -\int _0 ^t \langle \gamma (t') \rangle dt'
\right).
\end{equation}
All simulations shown in the figures correspond to averages over
1000 realizations.

We have plotted all results as functions of the non Gaussian noise
intensity $\xi _0 $. It is related to the (generating) white noise
intensity through  $\xi _0 = 2 D/(5 -3 q)$ \cite{we2}.

In what follows we show the results corresponding to the case
when the noise source is non Gaussian. In Fig. 1 we show the
amplitude of the absorption current ${\Delta \it J}(t)$ as a
function of the noise intensity $\xi _0$ for: (a) different $q's$
and fixed $\tau$ and observational time ($t$ ), (b) for three
different $\tau $ and fixed values of $q$ and $t$. The results
are in agreement with those found in the case of Gaussian white
noise. In the first case we see that the system response
increases when $q < 1$, and there is a shift of the maximum of
${\Delta \it J}(t)$ to larger noise values for increasing $q$. In
the second case the curves also show a shift of the maximum to
larger noise intensities as $\tau $ increases. Figure 2 shows the
resonant intensity noise values as a function of $\tau $ and for
different values of $q$. For $q < 1$ the maximum arises for lower
values of the noise intensity than for $q > 1$. In Fig. 3 we
depict the maximum of ${\Delta \it J}(t)$ as a function of $q$,
for two different observational times. Here we obtain one of the
main results of this work: the existence of new resonant-like
maximum as a function of the parameter $q$. This implies that we
can find an optimal value of $q$ ($\sim 0.5$, {\em corresponding
to a bounded and non Gaussian pdf}) yielding the largest system
response.

Finally, in Fig. 4 we show the value of ${\Delta \it J}(t)$  as a
function of $\tau $ for fixed $t$ and a large value of $\xi _0$.
The behaviour of $\Delta J$ is in agreement with the shift of the
curves shown in Fig. 1b. The inset shows the phase-shift between
the input and output signals, a result that is in agreement with
the main figure.

The present results show that the use of non Gaussian noises in
the simple trapping process defined in Eq. (\ref{model}) produces
significant changes in the system response when compared with the
Gaussian case. In particular we want to emphasize that we have
found a double resonance-like phenomenon indicating that, in
addition to an optimal noise intensity, there is an optimal $q$
value which yields the larger enhancement of the system response.
The remarkable fact is that it corresponds to $q < 1$ indicating
that this enhancement occurs for a non Gaussian and {\bf bounded}
distribution. Due to the evidences found in Ref. \cite{caltech},
such a result is of great relevance in a biological context.
Clearly, the present study corresponds to the analysis of a toy
model, more realistic ones will be the subject of further work. \\


{\bf Acknowledgments}

The authors want to thank to V. Gr\"unfeld for a critical reading
of the manuscript. Financial support from CONICET, Argentine,
is acknowledged.

\begin{figure}
\caption{Value of ${\Delta \it J}$ (amplitude of the oscillating
part of the absorption current) as a function of $\xi_{0}$ for a
given observational time ($t = 1140$). (a) different values of
$q$ (triangles $q = 0.5$, crosses $q = 1.0$, squares $q = 1.5$)
and a fixed value of $\tau $ ($\tau = 0.1$). (b)  different values
of $\tau $ (triangles $\tau = 0.01$, circles $\tau = 0.1$, squares
$\tau = 1.0$) and a fixed value of $q$ ($q = 0.5$).}
\end{figure}

\begin{figure}
\caption{Values of $\xi_{0}^{max}$, the noise intensity
corresponding to the maximum of ${\Delta \it J}$, as a function
of $\tau $ for fixed observational time ($t = 1140$) and
different values of $q$: circles $q = 0.5$, triangles $q = 1.5$.}
\end{figure}

\begin{figure}
\caption{Dependence of ${\Delta \it J}_{max}$, the value of
${\Delta \it J}$ at the maximum, as a function of $q$ for fixed
$\tau $ ($\tau = 0.1$) and different observational times: circles
$t = 633$, crosses $t = 1140$.}
\end{figure}

\begin{figure}
\caption{Dependence of ${\Delta \it J}$ as a function of $\tau $
for fixed values of $\xi _0$ ($\xi _0 = 4.7$), $q$ ($q = 0.5$) and
$t$ ($t = 1140$). The inset shows the phase shift $\phi$ as a
function of $\xi_{0}$ for different values of $\tau $: squares
$\tau = 0.01$, circles $\tau = 0.1$, triangles $\tau = 1.0$.}
\end{figure}

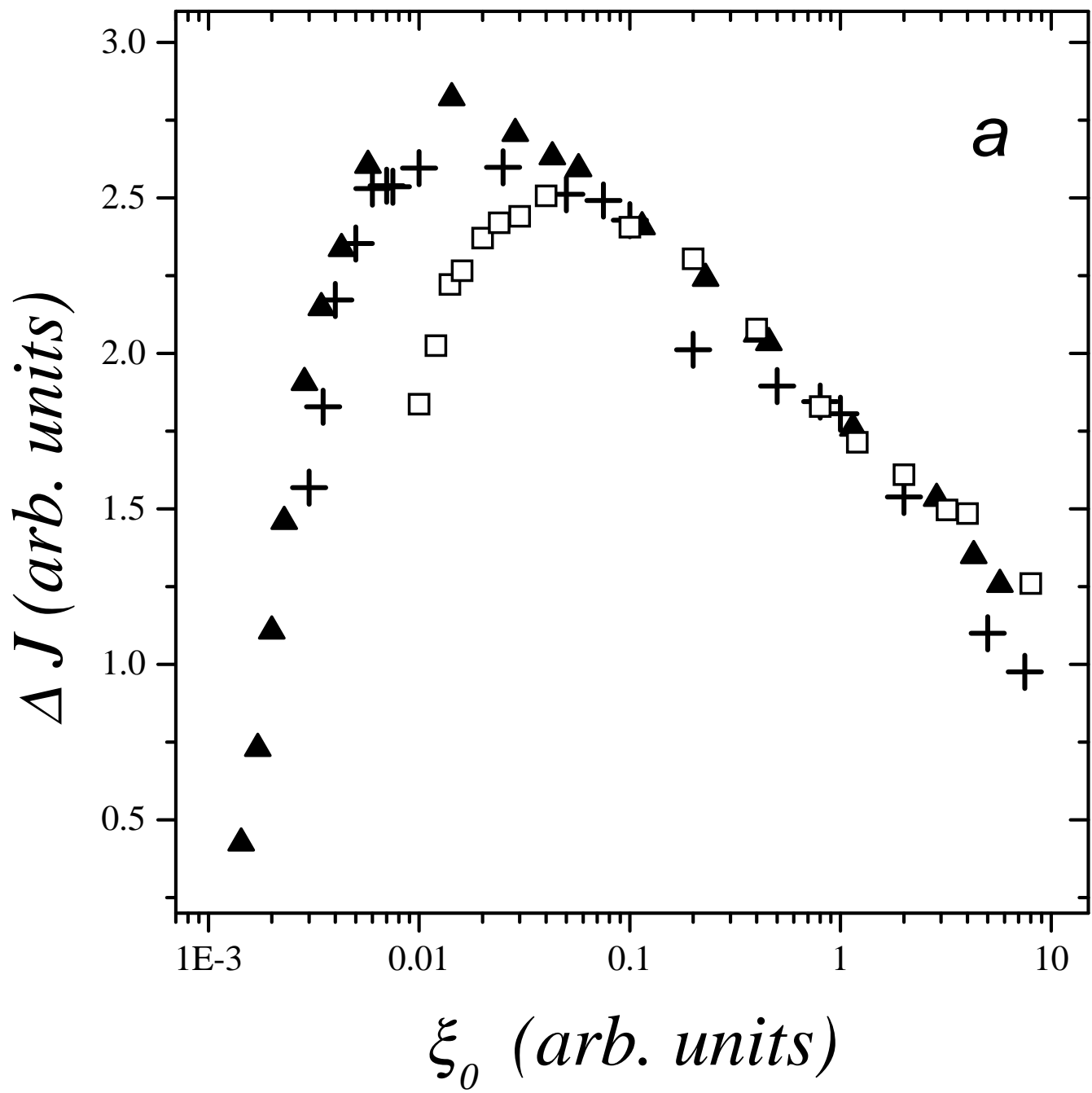
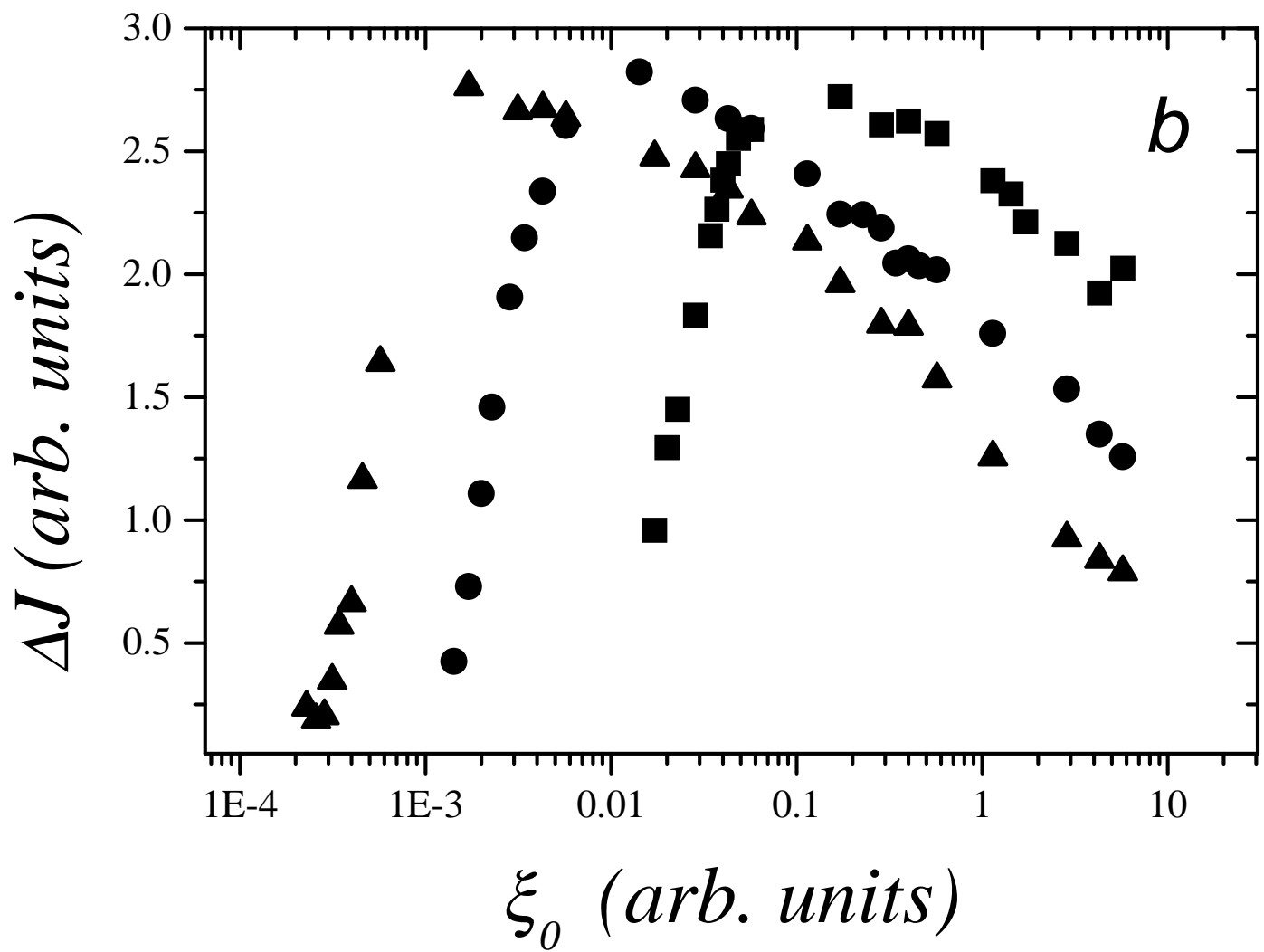
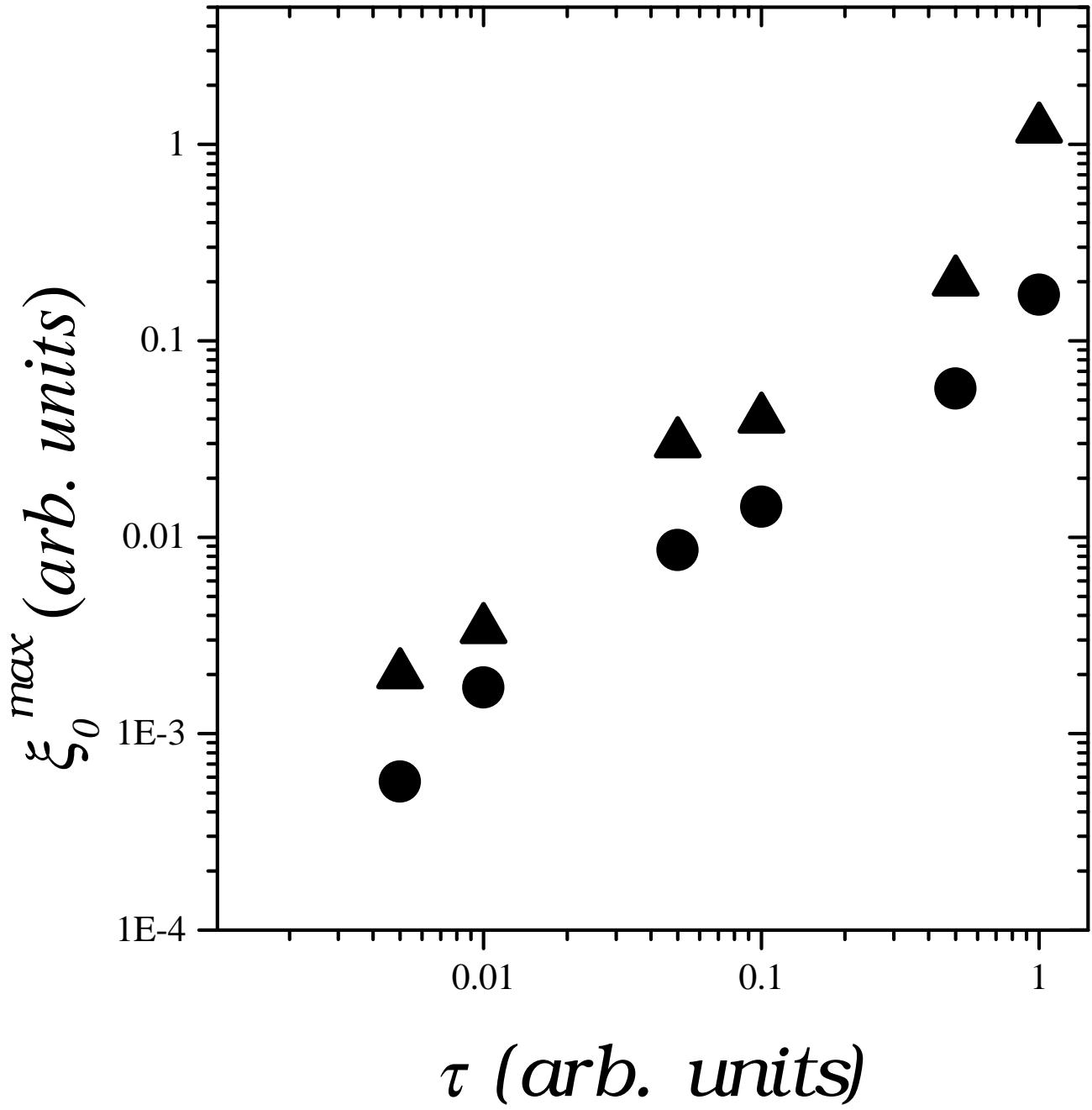
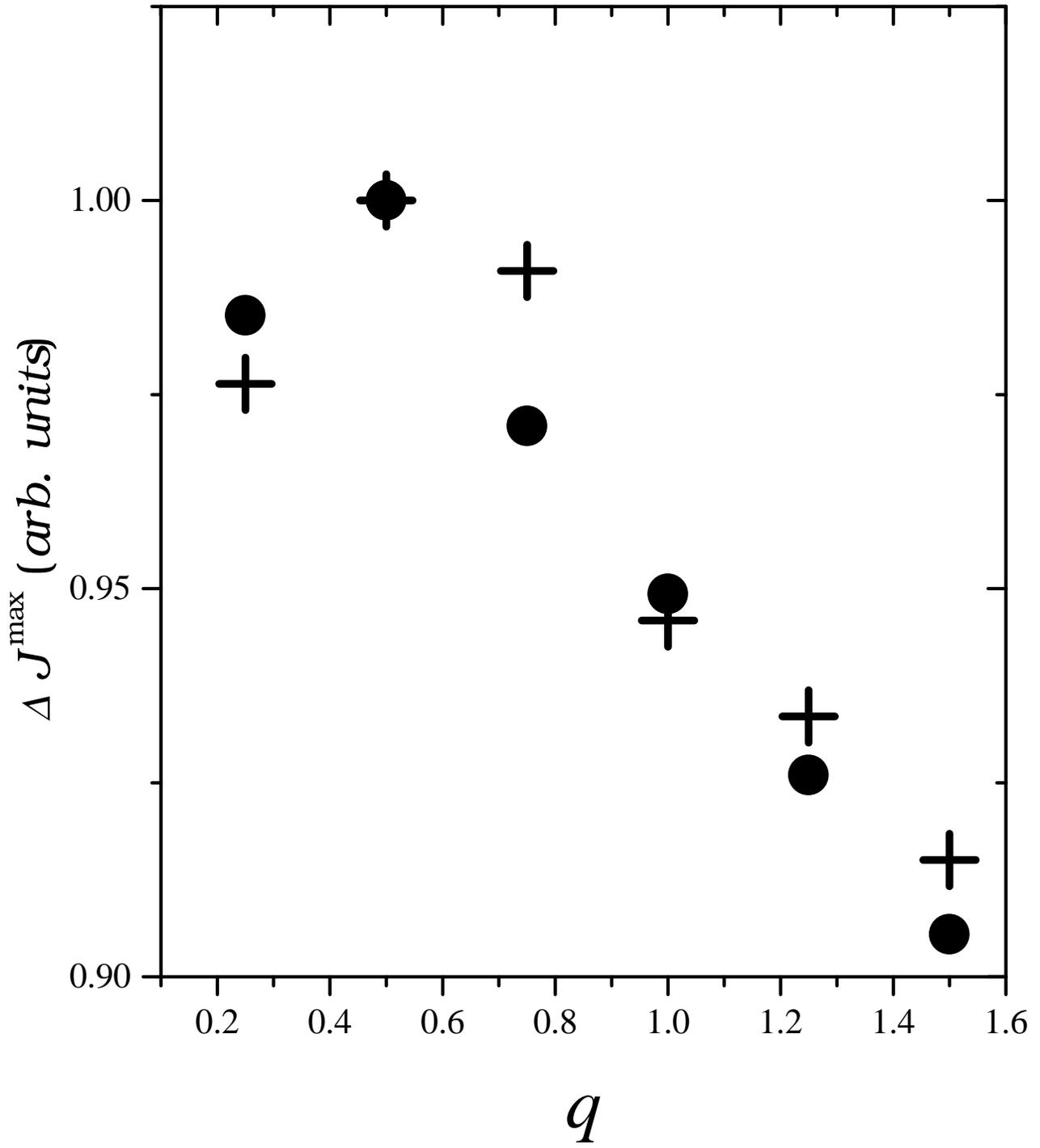
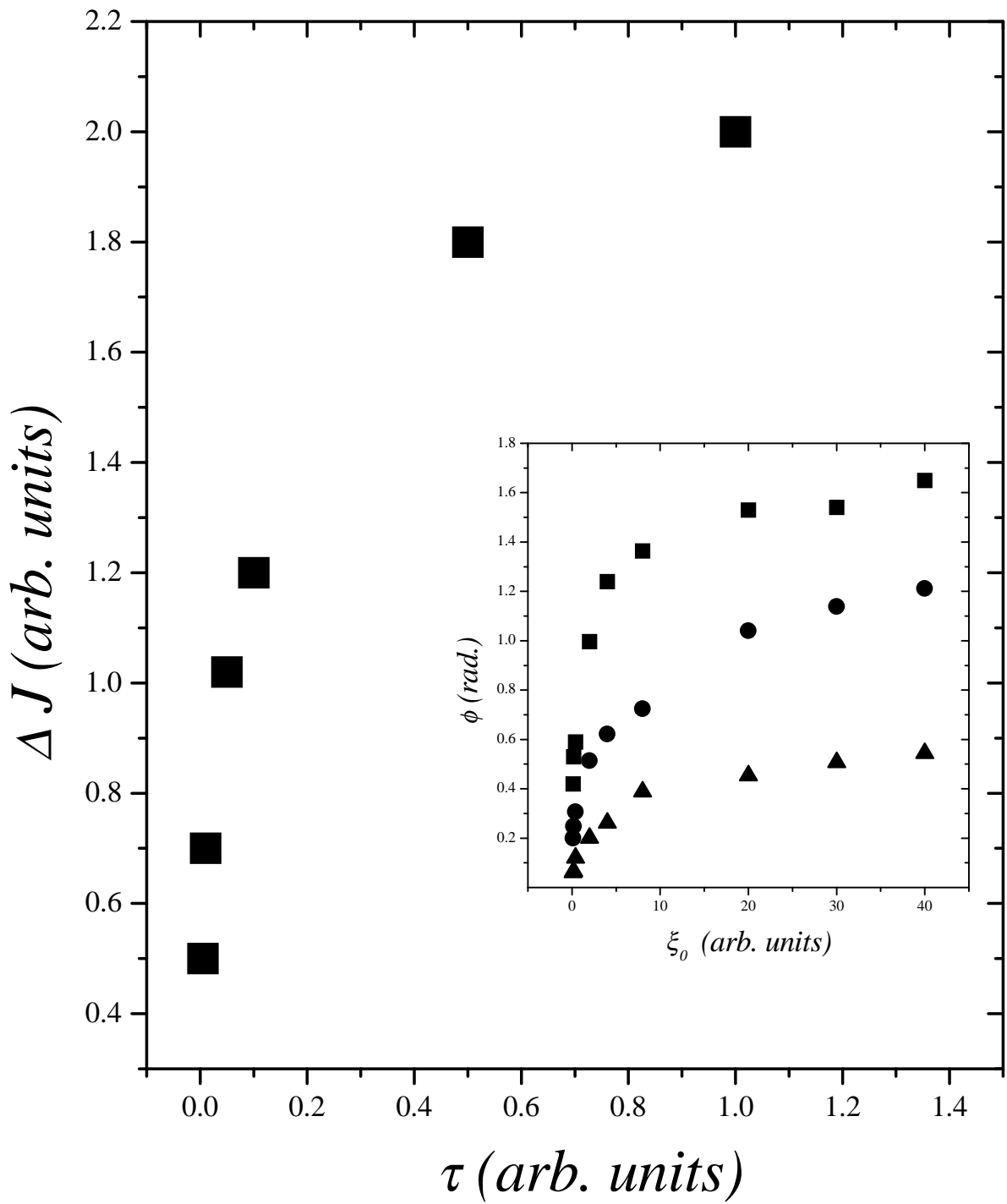
\end{document}